\documentclass[12pt]{iopart}
\usepackage{graphicx}
 
\begin{document}

\title[Continuous differential operators]{Continuous differential operators and a new interpretation of
                                       the charmonium spectrum}

\author{R Herrmann}

\address{GigaHedron, Farnweg 71, D-63225 Langen, Germany}
\ead{herrmann@gigahedron.de}
\begin{abstract}
The definition of the standard differential operator is extended from integer steps to arbitrary stepsize.
The classical, nonrelativistic Hamiltonian is quantized, using these new continuous operators.
The resulting Schroedinger type equation generates free particle solutions, which are confined in space.
The angular momentum eigenvalues are calculated algebraically. It is shown, that the charmonium spectrum
may be classified by the derived angular momentum eigenvalues for stepsize=2/3.   
\end{abstract}

\submitto{\JPG}

\section{Introduction of the continuous differential operator}

Since Newton{\cite{newton}}
 and Leibniz{\cite{leibniz}}
 introduced the concept of infinitesimal calculus,
differentiating a function with respect to the variable $x^i$ is a standard 
technique applied in all branches of physics. The differential operator $\partial_i$,

\begin{equation} 
\partial_i = \frac{\partial}{\partial x^i}
\end{equation} 
transforms like a vector, its contraction yields the Laplace-operator
\begin{equation} 
\Delta = \partial^i \partial_i
\end{equation} 
which is a second order derivative operator,   the essential contribution to establish a wave equation, 
which is the starting point to describe several kinds of wave phenomena.

Until now, the differential operator has only been used in integer steps. We want to extend the idea
of differentation to arbitrary, not necessarily integer steps.
A natural generalization is to search for an operator $D$ by setting
\begin{equation} 
D^m = \partial^n
\end{equation} 
where $m,n$ are integers. Formally, this is solved by extracting the $m$-th root
\begin{equation} 
D = \partial^{n/m} \qquad\qquad\qquad\qquad m,n \in N
\end{equation} 
or, even more general, we will introduce a continuous differential operator by
\begin{equation} 
D = \partial^p      \qquad\qquad\qquad\qquad\quad p \in R_+
\end{equation} 
with $p$ beeing a positive, real number.

In order to derive a specific representation for the operator $D$, we introduce a very simple
set of basis functions: 
\begin{equation} 
f(x) = x^\nu   \qquad\qquad\qquad\qquad\quad \nu \in R_+
\end{equation} 
Differentiation of these functions is well defined for the
integer case. It is just a multiplication with an appropriate factor and a linear reduction of the exponent.

This procedure can easily be extended to the continuous case.

The solution is inspired by a similar problem, which already has been solved in
mathematical history: the extension of the interpretation of the faculty function $n!$ to real numbers,
which resulted in  the famous Euler Gamma function $\Gamma(x+1)$, which is widely used in many different
areas of analysis.

Thus, the definition for the continuous derivative is:

\begin{equation}
D x^\nu =\partial^p x^\nu=
\cases{0                                           &  for $\nu = 0$ and $p>0$  \\
        \frac{\Gamma(\nu+1)}{\Gamma(\nu+1-p)}x^{\nu-p}   &  for $\nu,p > 0$, $\nu - p \ge 0$, $x \ge 0$ \\}
\end{equation}

The set of functions $\{x^\nu\}$ is sufficient to span a Hilbert space. 

Using the above definition,
we are able to quantize the Hamiltonian of a classical particle and solve the corresponding Schroedinger equation.

\section{Quantization of the classical Hamiltonian and free particle solutions}

We define the following set of conjugated operators
\begin{eqnarray}
P_\mu = \{  P_0,P_i\} &=& \{ i \hbar \partial_t,-i \hbar D_i \}\\
X_\mu = \{  X_0,X_i\} &=& \{ t, x^p_i / \Gamma(p+1)    \}
\end{eqnarray}
which satisfy the following commutator relations
\begin{equation}
[ X_i, X_j ] = 0, [ P_i, P_j ] = 0, [ X_i, P_j ] = -i \hbar \delta_{ij}
\end{equation}
With these operators, the classical, non relativistic Hamilton function $H_c$ 
\begin{equation}
H_c = \frac{p^2}{2m} +V(x^i) 
\end{equation}
is quantized. This yields the Hamiltonian H 
\begin{equation}
H = \frac{-\hbar^2}{2m} D^i D_i +V(X^i)
\end{equation}
Thus, a time dependent Schroedinger type equation for continuous differential 
operators results
\begin{equation}
\label{schroedinger}
H \Psi = \left( \frac{-\hbar^2}{2m} D^i D_i +V(X^i)\right) \Psi = i \hbar \partial_t \Psi
\end{equation}
For $p=1$ this reduces to the classical Schroedinger equation.

We will now present the free particle solutions for this equation. We can do this, since 
for $V(X^i) = 0$ the commutator $[P_\mu,H]$ 
vanishes and consequently, energy and momentum are conserved.

We extend the definition for the standard series expansion for the sine and cosine function
\begin{eqnarray}
\sin(p,x) &= \sum_{n=0}^\infty  (-1)^n x^{(2 n +1)p} / \Gamma((2n+1)p+1)  \\
\cos(p,x) &= \sum_{n=0}^\infty  (-1)^n x^{2 n p} / \Gamma(2np+1) \qquad \qquad x \in R_+ 
\end{eqnarray}
With this definition, the following relations hold
\begin{eqnarray}
D \sin(p,kx) &= k^p \cos(p,kx)  \\
D \cos(p,kx) &= -k^p \sin(p,kx) \qquad \qquad \qquad \quad k,x \in R_+
\end{eqnarray}
\begin{figure}
\begin{center}
\includegraphics[width=5in,height=3in]{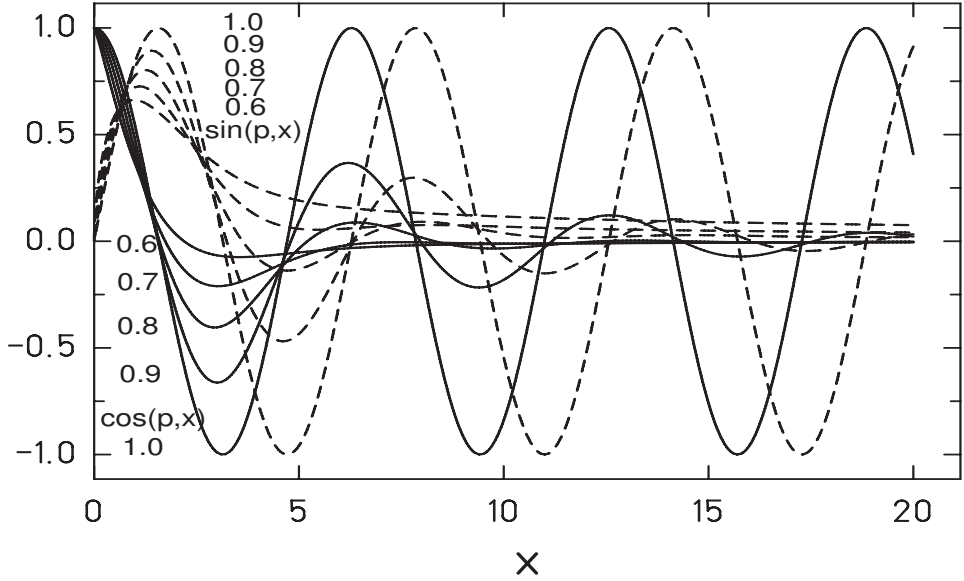}\\
\caption{\label{fig1}Free particle solutions for the continuous differential operator
Schroedinger type equation, for different values of $p$, $p=1,0.9,0.8,0.7,0.6$.
Solid lines are $\cos(p,x)$, dashed lines $\sin(p,x)$. For $p=1$ solutions
reduce to the standard $\cos(x)$ and $\sin(x)$ functions. For $p<1$ these functions are
increasingly located at $x=0$} 
\end{center}
\end{figure}
It follows from these relations, that this functions are the eigenfunctions of the free
Schroedinger type equation. In the stationary case we get the energy relation
\begin{equation}
E = \frac{\hbar^2}{2m} k^{2p}
\end{equation}

In figure 1 the functions $\sin(p,x)$ and $\cos(p,x)$ are plotted for different values
of $p$. While for $p=1$, these functions reduce to the known $\cos(x)$ and $\sin(x)$,
which are spread over the whole x-region,  
for $p<1$ these functions become more and more located at $x=0$ and oscillations are damped, a behaviour, which we
know e.g. from the Airy-functions. 

Let us make this observation more plausible: for any free particle solution $w(p)$ 
with $p=2/m$ and $m$ odd, which obeys the differential equation
\begin{equation}
D^2 w = -k^{2p} w
\end{equation}
differentiating $m-2$ times we obtain:
\begin{eqnarray}
D^{m-2}(D^2 w) &= D^{m-2}(-k^{2p} w)\\
D^m w &= -k^{2p} D^{m-2} w\\
\partial^2 w &= \pm k^{2p} k^{(m-2)p}(\frac{x^p}{\Gamma(1+p)} + R(p,x)) w , \quad \lim_{x \rightarrow 0} R(p,x) = 0\\
 &= \pm k^2 \frac{x^p}{\Gamma(1+p)} w + o(x^{3p}) 
\end{eqnarray}
This corresponds to an additional, impulse dependent potential with a leading linear term for similar 
solutions of the classical
Schroedinger equation ($p=1)$.

Thus, as the main result of our derivations so far, free particle solutions for $p\neq 1$ 
are localized in space.  Consequently, if there are particles, which show such an behaviour in nature,
a Schroedinger equation with continuous derivatives could be a useful tool for a description. 

In order to obtain more properties of the continuous differential operator Schroedinger equation, we will now
calculate the eigenvalues of the angular momentum operator.

\section{Classification of angular momentum eigenstates}
We define the generators of infinitesimal rotations in the $i,j$-plane
 ($i,j=1,...,3N$), $N$ is the
number of particles):
\begin{eqnarray}
L_{ij} &= X_iP_j - X_j P_i\\
       &= -i \hbar(\frac{x^p_i}{\Gamma(p+1)}D_j -\frac{x^p_j}{\Gamma(p+1)}D_i )
\end{eqnarray}
Since $[L_{ij},H] = 0$, angular momentum is conserved. 
Commutator relations for $L_{ij}$ are isomorph to a  $SO(3N)$ algebra:
\begin{equation}
[ L_{i   j  } ,    L_{m   n    } ] =
 i \hbar(
\delta_{i   m   } L_{j   n   } +
\delta_{j   n   } L_{i   m   } -
\delta_{i   n   } L_{j   m   } -
\delta_{j   m   } L_{ i  n   } )
\end{equation}
Consequently, we can proceed in a standard way {\cite{tb}}, 
by defining the Casimir-Operators
\begin{equation}
\Lambda_k^2=\frac{1}{2} \sum_{i,j}^{k} (L_{ij})^2 \qquad, \qquad k=2,...,3N
\end{equation}
which indeed fulfill relations 
$[\Lambda_{3N}^2, L_{ij} ] = 0$
and successively
$[\Lambda_k^2, \Lambda_{k'}^2 ] = 0$.
Their explicit form is given by
\begin{equation}
\Lambda_k^2 = -\hbar^2 (
X^i D_i + X^i X_i D^j D_j - \delta_i^i X^j
D_j - X^i X^j D_i D_j)
\end{equation}
Now we introduce a generalization of the homogenous Euler operator
for continuous differential operators 
\begin{eqnarray}
J_e^k(p) &= X^i D_i\\
          &= (x^i)^p/\Gamma(p+1) D_i
\end{eqnarray}
With the generalized Euler operator the Casimir-operators are:
\begin{equation}
\Lambda_k^2 = -\hbar^2 (
J_e^k + X^i X_i D^j D_j - k J_e^k - J_e^k(J_e^k-1))
\end{equation}
Now we define a Hilbert space $H_p$ of all homogenous polynoms of degree $l_k p$, which satisfy
the Laplace equation $D^iD_i f = 0$:
\begin{equation}
H_p = \{ f:f(\lambda x) = \lambda^{l_k p}f(x);\; D^i D_i f = 0\}
\end{equation}
On this Hilbert space, the generalized  Euler operator $J_e^k(p) $ is diagonal
and has the eigenvalues
\begin{equation}
l_k(p,n)  =
\cases{0                                           &  for $n=0$  \\
        \frac{\Gamma(np+1)}{\Gamma(p+1)\Gamma((n-1)p+1)}   &  for $n=1,2,3,...$\\}
\end{equation}
We want to emphasize, that these eigenvalues are different from the degree
of homogenity in the general case $p \neq 1$, or, in other words: only in the case of $p=1$
homogenity degree $n$ of the polynoms considered coincides with the eigenvalues of 
$J_e^k(p=1,n)$. 

Once the eigenvalues of the generalized Euler operator are known, 
the eigenvalues of the Casimir-operators $\Lambda_2, \Lambda_k^2$
are known, too:
\begin{eqnarray}
\Lambda_2 f & = & \hbar l_2(p,n)   f \\
\Lambda_k^2 f & = & \hbar^2 l_k(p,n)  (l_k(p,n)   + k - 2) f
\end{eqnarray}
with
\begin{equation}
l_k  \geq
l_{k-1}\geq...\geq l_2 \geq 0
\end{equation}
For the case of only one particle ($N=1$), we can introduce the quantum numbers
j and m, which now denote the j-th or m-th eigenvalue of the Euler operator.
The eigenfunctions are fully determined by these two quantum numbers $f = \mid jm>$ 

With the definitions $L_z = L_{12}$ and $J^2 = L_{12}^2 + L_{13}^2 + L_{23}^2 $ 
it follows
\begin{eqnarray}
\label{eqLz}
L_z \mid jm> & = & \hbar l_2(p,m)   \mid jm>     \qquad \qquad\qquad m=0,+1,+2,...,+j\\
\label{eqJ2}
J^2 \mid jm> & = & \hbar^2 l_3(p,j)  \left( l_3(p,j)+ 1 \right)  \mid jm> \qquad j=0,+1,+2,...
\end{eqnarray}
Please note the fact, that in the case $p \neq 1$ $L_z$ has no negative
eigenvalues. 

\begin{table}
\caption{\label{tabone}Eigenvalues in units of $\hbar$ for the angular momentum states of a single particle.
$n$ is a counter for the eigenvalues of the Euler operator, eigenvalues for $L_z$ and
$J^2$ are given for $p=1$ and $p=2/3$.} 

\begin{indented}
\lineup
\item[]\begin{tabular}{@{}*{5}{r}}
\br                              
$n$ & $L_z(p=1,m)$ & $L_z(p=2/3,m)$ & $J^2(p=1,j)$ & $J^2(p=2/3,j)$ \cr
\mr
0   &0         &0           &0                    & 0         \cr
1   &1         &1           &2                    & 2         \cr
2   &2         &1.460 998 49  &   6                 & 3.595 515 06\cr 
3   &3         &1.860 735 02  &  12                 & 5.323 069 84\cr 
4   &4         &2.222 222 22  &  20                 & 7.160 493 83\cr 
5   &5         &2.556 747 35  &  30                 & 9.093 704 37\cr 
6   &6         &2.870 848 32  &  42                 &11.112 618 39\cr 
\br
\end{tabular}
\end{indented}
\end{table}
In table 1 the first seven eigenvalues of $L_z$ and $J^2$ for a single particle are listed 
for $p=1$ and $p=2/3$. For $n>1$ and $p<1$ the eigenvalues of the Euler operator are not 
equally spaced any more, instead stepsize is strongly reduced. Since the Euler operator
eigenvalues contribute quadratically into the definition $J^2$, the energy of 
higher total angular momenta is reduced increasingly.

We have derived the full spectrum of the angular momentum operator for the
continuous differential operator Schroedinger type wave equation by use of 
standard algebraic methods. 

We expect additional information about the properties of this wave equation, if
we consider its linearized pendant. We present some results in the next 
section.  

\section{Results for the linearization of a second order differential equation}
Linearization of a second order wave equation was first considered by Dirac{\cite{dirac}}.
We propose a generalization of his approach, 
starting with a linear version of the wave equation, which is iterated
$m$ times, roughly:
\begin{equation}
(\gamma D)^m = \partial^2
\end{equation} 
This is formally solved by taking the $m$-th root:
\begin{equation}
\gamma D = (1)^{(1/m)} \partial^{2/m}
\end{equation} 
A new property is revealed via linearization: a phase factor, which may not be neglected.
If $m$ is an integer, $m$ different phase factors exist, which have to be incorporated 
within the definition of $\gamma$. Consequently, $\gamma$ may be interpreted as a matrix
with at least $m \times m $ dimensions.

The $\gamma$ matrices obey an extended Clifford algebra ($P$ means all permutations): 
\begin{equation}
\sum_{P} \prod_{i=1}^m \gamma_i = m! \delta_{i_1 i_2 .. i_m} 
\end{equation} 
This equation is fulfilled by a set of hermitean, traceless $m \times m$ matrices, which
built at least a subspace of SU(m).

Finally, we can distinguish two distinct parts of linarization. If spacelike components
are linearized ($D = D_i$), an additional SU(m) symmetry is introduced, which has 
been demonstrated for the case $m=2$ by Levy-Leblond by linearizing the 
ordinary Schroedinger equation to get 
an additional SU(2) symmetry{\cite{ll}}. For a single linearized 
timelike component, we 
obtain $m$ different energy values, which are interpreted as $m$ different particles.

From a mathematical point of view, a continuous differential operator may be
interesting for any value of p, a physicist may first concentrate on $p = 2/m$, because
there is a simple interpretation of the physical properties described by 
such a wave function.

We want to emphasize, that $p = 2/3$ seems to be the most promising candidate 
for a
physical interpretation:

It matches a triple iteration of a linear 
wave equation, which shows a direct connection to SU(3) and provides a description
for up to 3 particles simultaneously. Furthermore, as we have already shown, particles
are automatically confined in space. 

Summarizing all these facts, we assume, that the continuous differential operator
Schroedinger type wave equation with $p=2/3$ is an appropriate candidate for a 
non relativistic description of particles with quark-like properties.

\section{Interpretation of the Charmonium spectrum}
In the previous sections we have introduced the idea of continuous differential
operators and discussed some properties of the resulting non relativistic Schroedinger 
equation and its linearized pendant.


\begin{figure}
\begin{center}
\includegraphics[width=4in,height=5in]{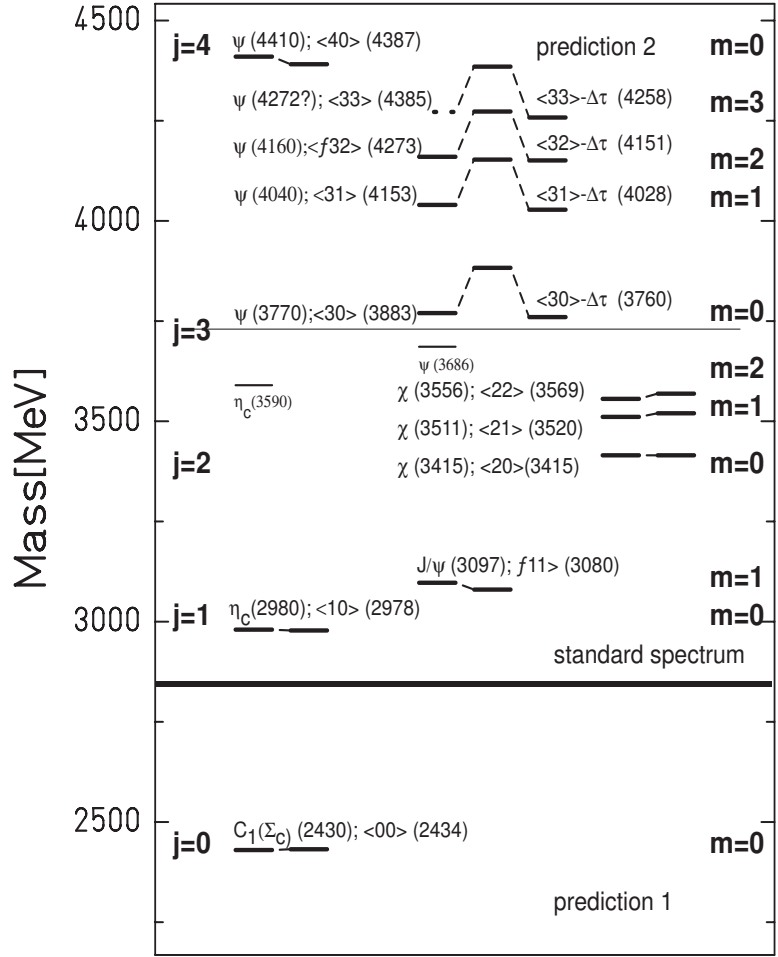}\\
\caption{\label{fig2}Charmonium spectrum. All observed particles are given with their name, experimental
mass, the proposed SO(3) conform quantum numbers and calculated mass 
according to  mass formulas ({\ref{massformula1}}),({\ref{massformula2}}). States are labeled with the derived SO(3)
quantum numbers, with j and m are the j-th and m-th eigenvalue of the generalized Euler operator. For $j=3$,
the influence of an additional constant correction term $\Delta \tau$ is illustrated.} 
\end{center}
\end{figure}

We have developed a new theoretical concept, which fulfills at least the following
three demands:
First, available experimental data will be reproduced with a reasonable accuracy.
Second, it will give new insights on underlying symmetries and properties
of the objects under consideration. 
Third, we will make predictions, which can be proven by experiment.

None of our results, presented so far, does require any information from QCD or a similar
theory. All our statements could have been made in the 1930ies already, even though they
would have been highly speculative. Today, we are in the comfortable position, that
there are enough experimental data, our predictions can be compared with.

A promising candidate is the charmonium spectrum{\cite{greiner}}. 

In the upper part of figure 2
we have displayed all experimentally observed charmonium-states with experimental masses, 
which are normally compared with results from a potential model, which tries to simulate
confinement and attraction by fitting a model potential{\cite{Ei75}},{\cite{Kr79}}.

We will assume, that this spectrum is a single particle spectrum for a 
particle, whose properties are described by the continuous differential operator 
Schroedinger equation({\ref{schroedinger}}) with $p=2/3$. We suppose, the system is rotating in a minimally
coupled field, which causes a constant magnetic field B. This leads to the following
 Hamiltonian H or mass formula
\begin{equation}
\label{massformula1}
H \mid jm> = m_0 c^2 + \alpha J^2 + B L_z \mid jm> 
\end{equation}
where $m_0 c^2$, $\alpha$ and B will be adjusted to the experimental data. The eigenfunctions 
are modified spherical harmonics with $D^i D_i \mid jm> =0$, the eigenvalues for $J^2$ and $L_z$
are given by ({\ref{eqLz}}),({\ref{eqJ2}}) and are listed in table 1.

In figure 2 we have sketched the resulting level scheme and the quantum numbers $<jm>$
for every particle. 

We have to prove, that this leads to correct results.

The first crucial test is the verification of the correct value of the non trivial $m=2$ quantum number
which corresponds to the $n=2$ eigenvalue of the generalized Euler operator. 
For the set of $\chi$-particles, we obtain:
\begin{equation}
L_z(j=2,m=2)_{exp}=\frac{\chi(3556)<22> -\chi(3415)<20>}{\chi(3511)<21> -\chi(3415)<20>} =  1.468
\end{equation}
This is an exact match with the theoretical value within the experimental errors, which are about $\pm 2[MeV]$.

Next we will determine the constants $m_0 c^2$ and $\alpha$. We choose the two lowest experimental states of the
standard charmonium spectrum, $\eta_c(2980)<10>$ and $\chi(3415)<20>$. 
This yields a set of equations
\begin{eqnarray}
m_0 c^2 + 2 \alpha           &= \eta_c(2980)<10>\\
m_0 c^2 + 3.59551506 \alpha  &= \chi(3415)<20> 
\end{eqnarray}
which determine $m_0 c^2 = 2434.72[MeV]$ and $\alpha = 272.64[MeV]$

Our level scheme predicts a particle with quantum numbers $<00>$, which is beyond the scope of 
charmonium potential models. According to our mass formula, it has a 
predicted mass of $m_0 c^2 = 2434.72[MeV]$. 

This is the second crucial test, since this is a low lying state, it should 
already have been observed.  Indeed, there exists an appropriate candidate, the 
$C_1(\Sigma_c)(2430)<00>$ particle {\cite{greiner}},{\cite{baltay}}.

This result supports our theory, since it is a second occurance of the $m=2$
eigenvalue of the generalized Euler operator.

In addition, it supports the idea, that the
energy levels of the spectrum are labeled correctly. 

Finally, the minimal difference of only $4[MeV]$, which
is mainly due to the uncertainty of the $\eta_c(2980)<10>$ experimental mass,
between predicted and experimental mass of the $C_1(\Sigma_c)(2430)<00>$ particle indicates, that the
assumed SO(3) symmetry is fulfilled exactly. 

As a last step, we need to determine the value of $B$. From the experimental data, we observe a slight 
dependence on $j$, which we will ignore for sake of simplicity. 
Instead, we do a least square fit for $B(j=1)$ and $B(j=2)$ and
obtain $B = 107[MeV]$.

Next step is to find higher eigenvalues of the generalized Euler operator. As a matter of fact, all 
candidates are positioned above $3.73[GeV]$, the threshold for $D\bar{D}$-meson production. 
From the point of view  of our theory, the particle properties change and we should actually make a new fit
for this particle with $H_D$  mass formula and new parameters. 
\begin{equation}
H_D \mid jm> = m_{0D} c^2 + \alpha_D J^2 + B_D L_z \mid jm> 
\end{equation}
For $B_D$ we obtain:
\begin{eqnarray}
\label{massformula2}
B_D &= \Psi(4040)<31> - \Psi(3770)<30> = 270[MeV]\\
B_D &= 2.5\, B
\end{eqnarray}
From a pessimistic point of view, this is just one more parameter, from an optimistic point of view
we could associate a correlation of B and R, the ratio of hadronic and mesonic cross sections, which
shows a similar behaviour{\cite{wolf}}.

We first prove the position of the $\Psi(4160)<32>$ state
\begin{equation}
L_z(j=3,m=2)_{exp}=\frac{\Psi(4160)<32> -\Psi(3770)<30>}{\Psi(4040)<31> -\Psi(3770)<30>} =  1.444
\end{equation}
This is in exact agreement with the theoretical value within the experimental errors, which are about $\pm 10[MeV]$ and the third observation of the $n=2$ 
eigenvalue of the generalized Euler operator.
This is an indication, that the proposed level scheme is still valid in the above region. 
As a consequence, we expect the $<33>$ state at
\begin{eqnarray}
\Psi(?)<33> &= \Psi(3770)<30> + 1.86073502 B_D \\
            &= 4272[MeV] \pm 15[MeV]
\end{eqnarray}
This state is missing in the experimental spectrum, so this is the second predicted particle or 
resonance, respectively in this area.

Consequently, we have only one experimental candidate for for $j=3$ and $j=4$ respectively. 
With the assumption $\alpha_D = \alpha$ and $m_{0D} c^2 =m_0 c^2 $ we obtain the theoretical values
\begin{eqnarray}
\Psi(3770)<30>_{th} &= 3886[MeV]\\
\Psi(4410)<40>_{th} &= 4387[MeV]
\end{eqnarray}
For $j=3$ the calculated mass differs by $116[MeV]$ from the experimental value. In other words, 
the $j=3$ eigenvalue of the generalized Euler operator may be deduced from experiment as $1.77$ 
instead of $1.86$ from the theory. This is 
a deviation of about $5\%$, which is close, but compared with the previous results not precise.

On the other hand, the theoretical $\Psi(4410)<40>$ mass  
matches exactly with the experimental value within the experimental errors.

This indicates, that the particles for $j=3$, observed in experiment, carry an additional property, which 
reduces the mass by the amount of e.g. a pion. Of course,
if we add an additional $(\Delta\tau) \delta_{j3}$ term to the proposed mass formula, we can shift these levels 
by the necessary amount. Figure 2 shows the theoretical masses for $j=3$ with such an correction term switched off and on.

Summarizing these results, the charmonium spectrum reveals an underlying SO(3) symmetry, which agrees
with the predictions of our theory in the case of $p=2/3$. The eigenvalues of the generalized
Euler operator conform within experimental errors with experimental data for $n=0,1,2,4$, for $n=3$ there is 
a deviation of $5\%$.
Two new particles have been predicted.
With only 4 parameters (including $\Delta \tau$) the massformulas ({\ref{massformula1}}),({\ref{massformula2}}) reproduce
the charmonium spectrum within a range of $2 GeV$.

\section{Conclusion}
We have defined a continuous differential operator for arbitrary stepsize $p$. A Schroedinger type wave equation,
derived by quantization of the classical Hamiltonian, bases on these operators, generates  free particle solutions, 
which are
confined to a certain region of space. The multiplets of the generalized angular momentum operator have been  classified
acoording to the SO(3) scheme, the spectrum of the Casimir-Operators has been calculated analytically. We have also
shown, that for $p=2/3$, corresponding to a  triple iterated linearized wave function an inherent SU(3) 
symmetry is apparent. 

From a detailed discussion of the charmonium spectrum we conclude, that the spectrum may be
understood within the framework of our theory.

Two new particles have been 
predicted, one of them already confirmed by experiment, the other being a resonance at $4272[MeV] \pm 15[MeV]$.

The results indicate, that continous differential operator wave  equations may play an important role
for our understanding of phenomena in the area of particles with quark-like properties, e.g. confinement.

Up to now, we have only derived a non relativistic wave equation. A next step will include 
relativistic effects too.


\section*{References}
\begin{thebibliography}{10}
\bibitem{newton} Newton I 1669 {\it De analysi per aequitiones numero terminorum infinitas}, manuscript
\bibitem{leibniz} Leibniz  G F Nov 11, 1675 {\it Methodi tangentium inversae exempla}, manuscript
\bibitem{tb} Louck J D and Galbraith H W 1972 Rev.Mod.Phys. {\bf 44(3)}, 540
\bibitem{dirac} Dirac P A M 1928 Proc.Roy.Soc. (London) {\bf A117}, 610
\bibitem{ll} Levy-Leblond J M 1967 Comm.Math.Phys. {\bf 6}, 286
\bibitem{greiner} Greiner W and M\"uller B 2001 {\it Quantum Mechanics, Symmetries} Springer Berlin, New York
\bibitem{Ei75} Eichten E, Gottfried K, Kinoshita T, Kogut J, Lane K D and Yan T M 1975 Phys.Rev.Lett. {\bf 34}, 369
and 1976 Phys.Rev.Lett. {\bf 36}, 500
\bibitem{Kr79} Krammer M and Krasemann H  1979 {\it Quarkonia in Quarks and Leptons} Acta Physica Autriaca,
Suppl. XXI, 259
\bibitem{baltay} Baltay C et. al. 1979 Phys. Rev. Lett. {\bf 42}, 1721
\bibitem{wolf} Wolf G 1980 {\it Selected Topics on $e^+e^-$-Physics} DESY 80/13
\end{thebibliography}
\end{document}